\documentclass[aps,prb,preprint,amsmath,amssymb,]{revtex4-2}
\usepackage{graphicx}
\usepackage[percent]{overpic}
\usepackage{subfigure}
\usepackage{physics}
\usepackage{spin}
\usepackage{xcolor}
\usepackage{ulem}
\usepackage{braket}
\usepackage{bm}
\usepackage{natbib}
\usepackage{array}

\newif\iffigure
\figuretrue 

\begin{document}

\title{Antiparallel spin polarization and spin current \\ 
induced by thermal current and locally-broken inversion symmetry \\
in a double-quantum-well structure}

\author{
Yuta Suzuki$^{*,1,2}$(Yuuta.Suzuki@yokogawa.com) \\
Yuma Kitagawa$^{1,2}$
Shin-ichiro Tezuka$^{2}$
and Hiroshi Akera$^{3}$}

\affiliation{
$^{1}$Division of Applied Physics, Graduate School of Engineering, Hokkaido University, Sapporo, Hokkaido, 060-8628, Japan\\
$^{2}$Sensing Research \& Development Department, Innovation Center, Marketing Headquarters, Yokogawa Electric Corporation, Tokyo, 180-8750, Japan\\
$^{3}$Division of Applied Physics, Faculty of Engineering, Hokkaido University, Sapporo, Hokkaido, 060-8628, Japan
}

\date{\today}

\begin{abstract}
Generating a nonequilibrium spin polarization with a driving force has been first realized 
by the electric current in a system with broken inversion symmetry 
and extended to that induced by the thermal current 
and that appearing in an inversion-symmetric system with locally-broken inversion symmetry. 
This paper theoretically explores the spin polarization generated 
by the thermal current and the locally-broken inversion symmetry 
in a symmetric double-quantum-well structure (DQWS).  
This thermally-induced spin polarization (TISP) 
appears in the antiparallel configuration with the TISP of two wells in opposite directions. 
The calculation using the Boltzmann equation in the relaxation-time approximation 
under the condition of zero charge current  
shows that the local TISP exhibits the maximum 
at a finite Rashba spin-orbit interaction when the electron density is fixed. 
This is because the local TISP in the DQWS is enhanced at the chemical potential near the bottom of the first-excited subband. 
This enhancement also occurs in a single quantum well with globally-broken inversion symmetry. 
Another finding is that the maximum of the local TISP appears at a nonzero interwell coupling. 
The spin current by the diffusion of the local TISP into an adjacent electrode is also calculated.  
\end{abstract}

\maketitle

\section{Introduction}

Current-induced spin polarization (CISP) 
\cite{Edelstein1990, Kalevich1990, Silov2004, Kato2004CISP, Yang2006, Trushin2007}
is generated by breaking the time-reversal symmetry with current 
in a system with broken inversion symmetry through the action of the spin-orbit interaction (SOI). 
The CISP 
on surfaces of topological insulators \cite{Mellnik2014, Kondou2016, Kondou2018} 
and in two-dimensional electron systems (2DES)  
\cite{Jungfleisch2016, Shao2016, Wang2017, Liu2020sot, Karube2020}
has been demonstrated to create the spin diffusion current 
which can be used for switching the magnetization in ferromagnetic memory. 

Thermally-induced spin polarization (TISP) has also been 
studied theoretically \cite{Wang2010, Dyrdal2013, Tolle2014, Xiao2016, Dyrdal2018} 
in the 2DES with the Rashba SOI \cite{Ohkawa1974, Bychkov1984a, Bychkov1984b, Faniel2011}. 
In the TISP the time-reversal symmetry is broken by the thermal current, 
which is the flow of electronic excitations from a lower to a higher eigenstate. 
Therefore the TISP is large 
when the difference between spin polarizations of the two eigenstates is large 
and signs of differences are the same for all excitations. 
We expect that such requirements can be met by adjusting system parameters. 

Locally-broken inversion symmetry in a system with the global inversion symmetry 
\cite{Yanasezigzag2014, Zhang2014, Zelezny2014, Riley2014, Wadley2016, Gehlmann2016, Santos-Cottin2016, Wu2017, Yao2017, Razzoli2017, Watanabe2018, Cheng2018, Yuan2019, Suzuki2023, Kitagawa2023, Kato2023} 
has also been used to create the CISP in the antiparallel configuration.  
Such antiparallel CISP is a nonequilibrium analog of the equilibrium antiferromagnetism.  
Consider an inversion-symmetric system consisting of two sublattices or two layers. 
If the inversion symmetry in each sublattice (each layer) is broken 
to produce a local effective magnetic field acting on spin, 
the sum of the spin density over a pair of spin-degenerate eigenstates is locally nonzero \cite{Zhang2014}, 
although it vanishes globally when integrated over the two sublattices (layers). 
By breaking the time reversal symmetry, 
this antiparallel spin density gives rise to the antiparallel CISP \cite{Yanasezigzag2014, Zelezny2014, Suzuki2023, Kitagawa2023, Kato2023}.
This antiparallel CISP has been demonstrated to 
reverse the sublattice magnetization of antiferromagnet \cite{Wadley2016, Watanabe2018} 
in antiferromagnetic memory \cite{Jungwirth2016}. 
The antiparallel CISP can also be used to generate the spin current by a selective coupling of an electrode to one sublattice (layer)  
as theoretically demonstrated in a double-quantum-well structure (DQWS) \cite{Suzuki2023} 
and in a buckled atomic layer \cite{Kitagawa2023}. 
The magnitude of the spin current has been shown to be comparable to 
that generated by globally-broken inversion symmetry \cite{Suzuki2023, Kitagawa2023}. 

\begin{table}[h]
\caption{Current-induced spin polarization and thermally-induced spin polarization.}
\vskip 0.2cm
	\label{table:research_position}
	\centering
	\begin{tabular}{|wc{3.5cm}|wc{3.5cm}|wc{3.5cm}|}
		\hline inversion symmetry
			&	globally broken &	locally broken \\
		\hline
			CISP 	
			&	Refs.\cite{Edelstein1990, Kalevich1990, Silov2004, Kato2004CISP, Yang2006, Trushin2007}  
			&	Refs.\cite{Yanasezigzag2014, Zelezny2014, Suzuki2023, Kitagawa2023, Kato2023} 	\\
		\hline
			TISP 	
			&	Refs.\cite{Wang2010, Dyrdal2013, Tolle2014, Xiao2016, Dyrdal2018}  	
			&	this study\\
		\hline
	\end{tabular}
\end{table}
	
In this study we employ both the thermal current and the locally-broken inversion symmetry 
to produce the spin polarization 
(\tablename\ref{table:research_position} shows the position of this study in the field).  
We choose the DQWS 
\cite{Koga2002spin_filter, Jin_Li2007, Bernardes2007, Calsaverini2008, Akabori2012, Hernandez2013, Souma2015, Khaetskii2017, Hayashida2020} 
as the simplest system with the locally-broken inversion symmetry. 
In the DQWS, which is symmetric with respect to the plane between two wells,  
the Rashba SOI gives the antiparallel effective magnetic field. 
We calculate the thermally-induced antiparallel spin polarization (antiparallel TISP) 
by using the Boltzmann equation in the relaxation-time approximation 
under the condition of zero charge current.  
We plot the local TISP ($\lTISP$) in one well 
as a function of the strength of the Rashba SOI and that of the interwell coupling 
with an aim to search optimum values of these system parameters 
which maximize the magnitude of the $\lTISP$. 
We find that the $\lTISP$ exhibits the maximum at a finite Rashba SOI and a nonzero interwell coupling 
when the electron density is fixed. 

The organization of this paper is as follows. 
Section \ref{sec:General_model} presents a general scheme 
for generating the spin current using the $\lTISP$ in a system with the locally-broken inversion symmetry.  
Section \ref{sec:DQWS_Model} describes the Hamiltonian of the DQWS and that of the electrode 
with their eigenstates.  
In Section \ref{sec:BoltzmannEquation} we employ the Boltzmann equation in the relaxation-time approximation  
to derive the distribution function of the DQWS 
with in-plane gradients of the temperature and the electrochemical potential. 
Section \ref{sec:Calculated_Results} presents calculated results of the $\lTISP$ 
under the condition of zero charge current. 
In addition this section derives the formula for the spin current flowing to the electrode, 
by introducing the Hamiltonian describing the tunneling between the DQWS and the electrode, 
and shows that the spin current is proportional to the $\lTISP$ 
under an assumption presented there with respect to the electron tunneling. 
Conclusions are given in Section \ref{sec:Conclusions}.

\section{Model and Hamiltonian}
\label{sec:Model_Hamiltonian}
\subsection{General Model} 
\label{sec:General_model}

Figure \ref{fig:model}(a) presents a general model for the thermal generation of the spin current from a system with the inversion symmetry, 
while Figure \ref{fig:model}(b) and (c) show two example systems, 
the DQWS and the group-IV atomic layer, respectively.  
The temperature difference between reservoir 1 and 2, $T_1 - T_2 \ (>0)$, 
is introduced to produce the electron-temperature gradient $\vnabla\Te$, 
which thermally induces the antiparallel spin polarization in A and B 
[layers L and R in Figure \ref{fig:model}(b) and sublattices A and B in Figure \ref{fig:model}(c)]. 
This thermally-induced local spin polarization ($\lTISP$) in each of A and B is then 
selectively extracted into electrodes A and B, respectively. 
We impose the condition of the vanishing charge current \cite{Ashcroft_Mermin1976solid},  
which produces the electrochemical-potential gradient $\vnabla\muec$. 
We describe the eigenstate of the $\lTISP$ generator as $\Ket{\nu\vk}$ 
with the wave vector $\vk=(k_x, k_y)$ and the band index $\nu$. 
In the electrodes we neglect the SOI and choose the eigenstate 
to be the eigenvector of $\hsigy$ the $y$ component of the Pauli spin operator,  
which satisfies $\hat{\sig}_y\ket{\sigy} = \sigy\ket{\sigy}$ with $\sigy = \pm1$. 
Here we take the $y$ axis in the direction of the $\lTISP$. 
Then the eigenstate of the electrode is expressed as $\Ket{\xi\eta\sigy}$, 
where $\xi=\EA, \EB$ and $\eta$ represents other quantum numbers.

\iffigure
\begin{figure}
\centering
\begin{overpic}[scale=.45, ]{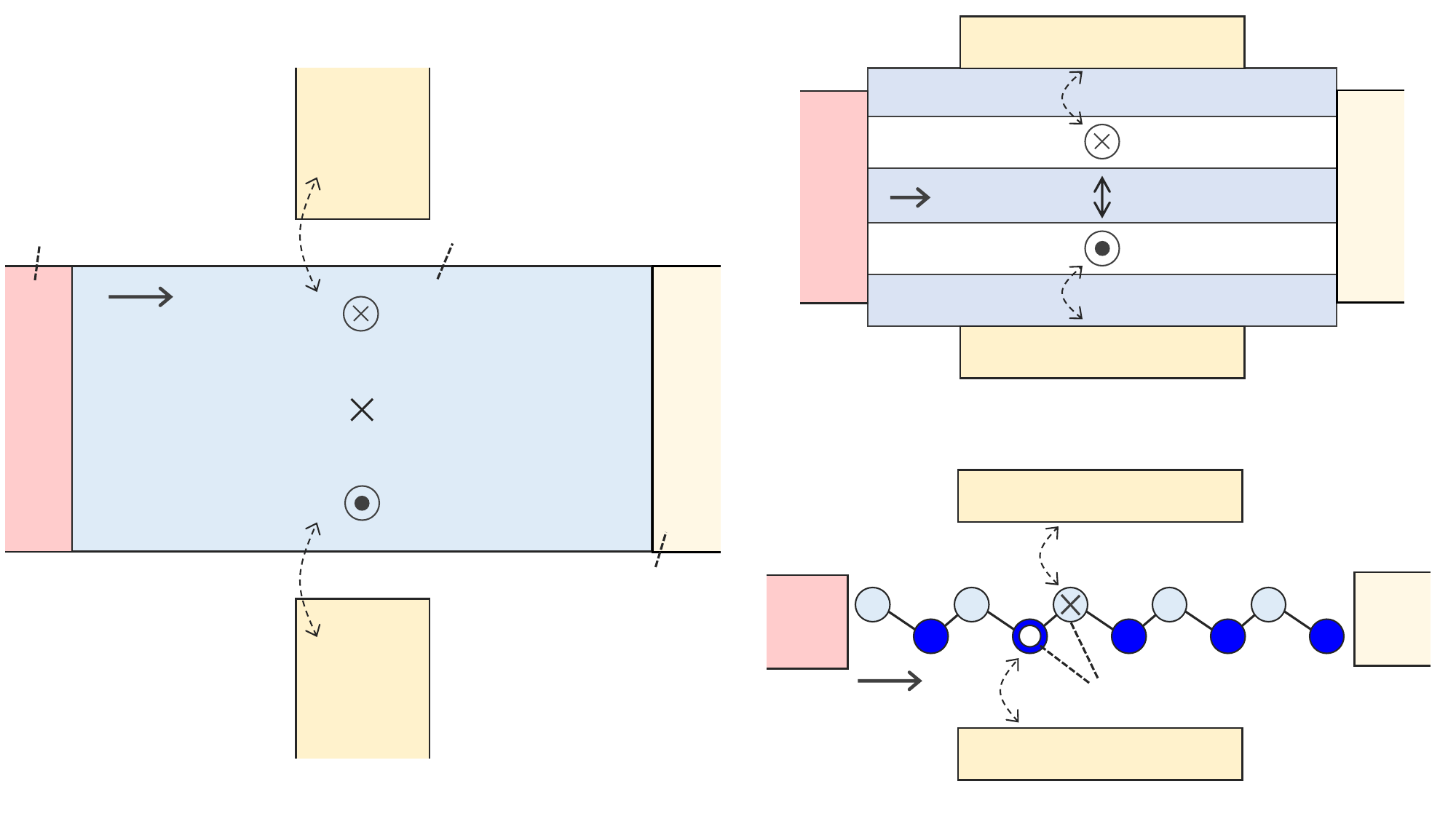}
	\put(	1,	54	){(a)}
	\put(	30,	44.5	){Electrode A}
	\put(	30,	9.5	){Electrode B}
	\put(	27,	27.5	){Inversion center}
	\put(	27,	33.5	){$\lTISP$}
	\put(	27,	21	){$\lTISP$}
	\put(	20.5,	33.8	){A}
	\put(	20.5,	21	){B}
	\put(	1,	40	){Reservoir 1}
	\put(	1.3,	29	){$T_1$}
	\put(	0.8,	25	){$\muecone$}
	\put(	37,	15	){Reservoir 2}
	\put(	46,	29	){$T_2$}
	\put(	45.5,	25	){$\muectwo$}
	\put(	6,	33	){$-\bm{\nabla}T_e$}
	\put(	30,	40	){$\lTISP$ generator}
	\put(	16,	15.5	){$H_\mathrm{T}$}
	\put(	16,	39	){$H_\mathrm{T}$}
	\put(	10,	27.5	){$\Ket{\nu\vk}$}
	\put(	22,	45	){$\Ket{\eta\sigy} $}
	\put(	22,	10	){$\Ket{\eta\sigy} $}
	\put(	53,	54	){(b)}
	\put(	78,	45.5	){$\lTISP$}
	\put(	78,	38.1	){$\lTISP$}
	\put(	66,	45.5	){$\alpha$}
	\put(	64,	38	){$-\alpha$}
	\put(	64.5,	41.7	){$-\bm{\nabla}T_e$}
	\put(	68,	49	){$H_\mathrm{T}$}
	\put(	68,	34.8	){$H_\mathrm{T}$}
	\put(	77,	42.2	){$\DSAS$}
	\put(	60,	45.5	){L}
	\put(	60,	38 	){R}
	\put(	69,	52.2	){Electrode L}
	\put(	69,	31.1	){Electrode R}
	\put(	52.2,	37	){\rotatebox{90}{Reservoir 1}}
	\put(	55.9,	44	){$T_1$}
	\put(	55.1,	41	){$\muecone$}
	\put(	97.3,	37	){\rotatebox{90}{Reservoir 2}}
	\put(	92.9,	44	){$T_2$}
	\put(	92.3,	41	){$\muectwo$}
	\put(	53,	24	){(c)}
	\put(	75.5,	7.5	){$\lTISP$}
	\put(	56.5,	6.3	){$-\bm{\nabla}T_e$}
	\put(	67.5,	17.5	){$H_\mathrm{T}$}
	\put(	65,	8	){$H_\mathrm{T}$}
	\put(	81,	16.5	){A site}
	\put(	85,	9 	){B site}
	\put(	69,	21.2	){Electrode A}
	\put(	69,	3.6	){Electrode B}
	\put(	52,	18	){Reservoir 1}
	\put(	53.5,	13.8	){$T_1$}
	\put(	53.2,	11.2	){$\muecone$}
	\put(	90,	18	){Reservoir 2}
	\put(	95,	13.8	){$T_2$}
	\put(	94.8,	11.2	){$\muectwo$}
\end{overpic}
\caption{\label{fig:model}
	(a) General model for the generation of the spin current by $\lTISP$.
	(b) DQWS model.
	(c) Atomic-layer model.
}
\end{figure}
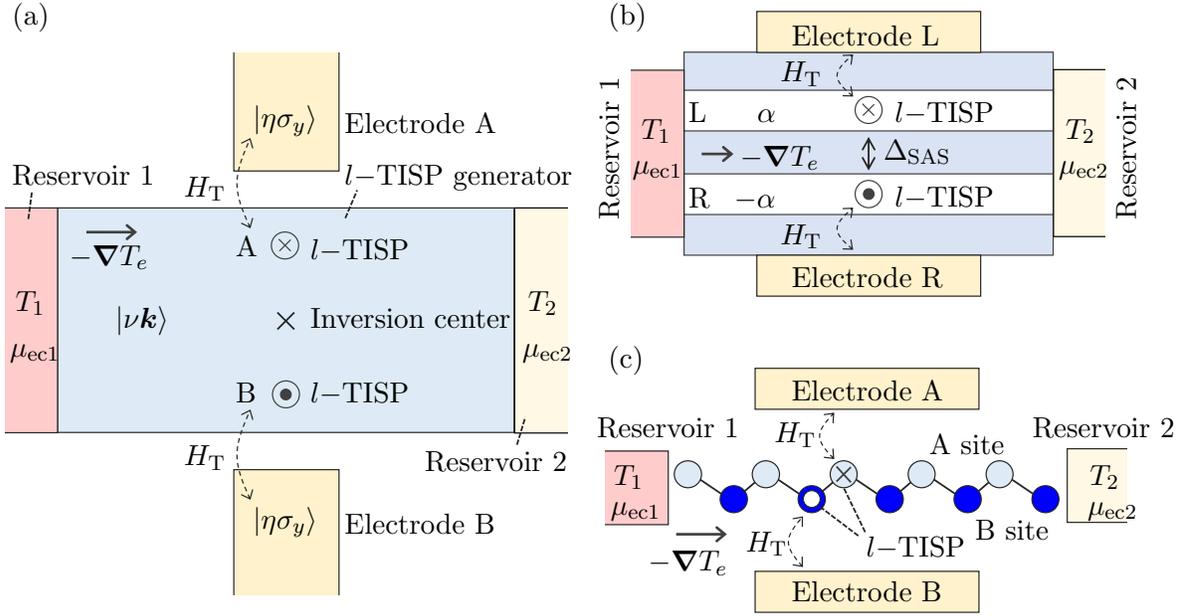
\fi

We describe the coupling between the $\lTISP$ generator and each electrode by the tunneling Hamiltonian $H_\mathrm{T}$. 
Then the Hamiltonian of our general model in \figurename\ref{fig:model}(a) is expressed by
\begin{alignat}{99}
	H =  H_0 + H_\mathrm{El} + H_\mathrm{T},
\end{alignat}
where $H_0$ and $H_\mathrm{El}$ are the Hamiltonian of the $\lTISP$ generator 
and that of the electrodes, respectively. 
Their eigenvectors satisfy 
\begin{alignat}{99}
	H_0 \Ket{\nu\vk} = \epsi_{\nu\vk} \Ket{\nu\vk}, \quad
	H_\mathrm{El} \Ket{\xi\eta\sigy} = \epsi_{\eta} \Ket{\xi\eta\sigy},\ \ \xi=\textrm{A, B}
	\label{eq:GeneralSchroedinger_Eq}
\end{alignat}
where $\epsi_{\nu\vk}$ and $\epsi_{\eta}$ are the corresponding eigenvalues. 

\iffigure
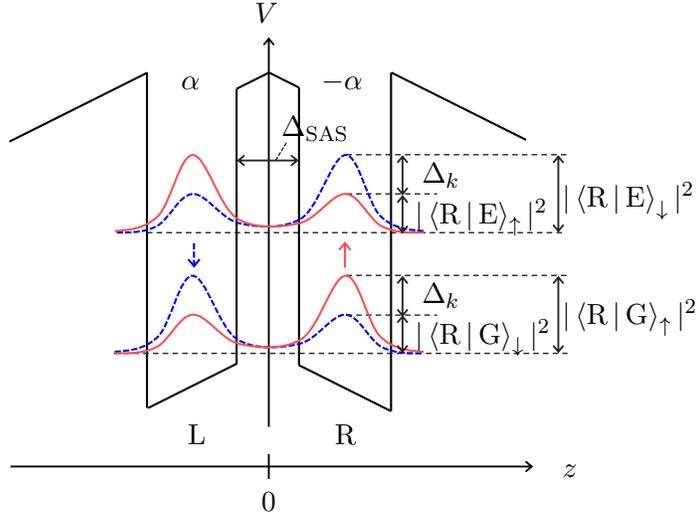
\begin{figure}
\centering
\begin{overpic}[scale=.35, ]{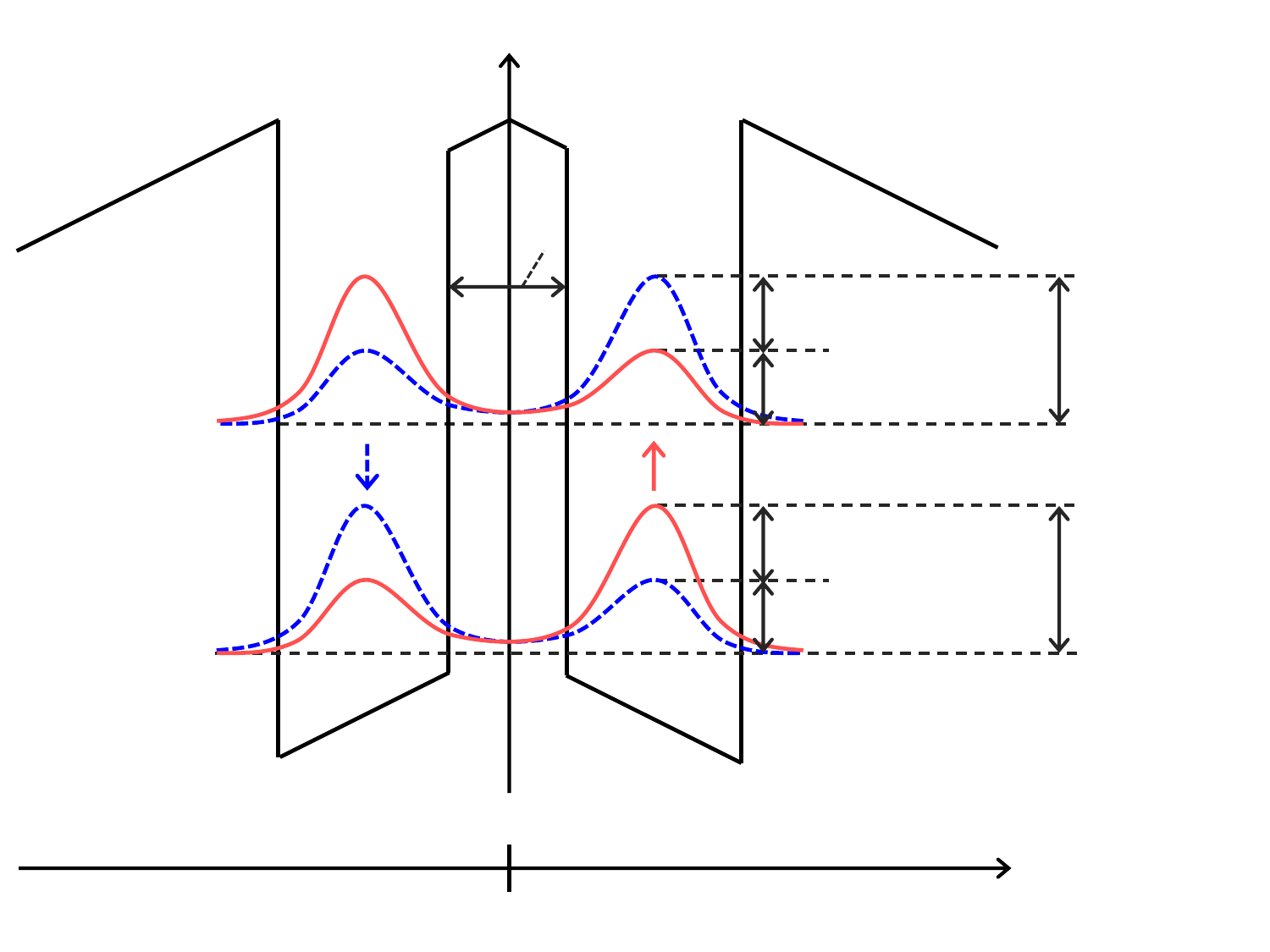}
	\put(	28,	10	){L}
	\put(	50,	10	){R}
	\put(	38,	73	){$V$}
	\put(	84,	5	){$z$}
	\put(	42,	56	){$\DSAS$}
	\put(	39,	0	){0}
	\put(	27,	63	){$\alpha$}
	\put(	48,	63	){$-\alpha$}
	\put(	63,	49	){$\Delta_k$}
	\put(	62,	43	){$|\Braket{\WR|\WE}_{\uparrow}|^2$}
	\put(	84,	45.5	){$|\Braket{\WR|\WE}_{\downarrow}|^2$}
	\put(	63,	31	){$\Delta_{k}$}
	\put(	62,	25	){$|\Braket{\WR|\WG}_{\downarrow}|^2$}
	\put(	84,	28	){$|\Braket{\WR|\WG}_{\uparrow}|^2$}
\end{overpic}
\caption{\label{fig:dqw_energy_diagram}
Potential $V$ and the local spin density in each eigenstate of the DQWS [\figurename\ref{fig:model}(b)]. 
Red solid and blue dashed lines 
represent the schematic distribution of spin-up and spin-down states, respectively,  
with wave number $k$ in the ground subband G and the first excited subband E. 
The direction of spin is parallel or antiparallel to the effective magnetic field.  
$\Delta_k$ is the magnitude of the local spin polarization in each well.
}
\end{figure}
\fi

\subsection{DQWS as the simplest example}
\label{sec:DQWS_Model}
We study the $\lTISP$ in the DQWS, 
as the simplest example of inversion-symmetric structure, in the following. 
The DQWS [Figure \ref{fig:model}(b)] 
consists of two wells, L and R, 
with the Rashba spin-orbit coefficients, $\alpha(>0)$ and $-\alpha$, respectively. 
The strength of the coupling between L and R wells is denoted by $\DSAS$. 
Wells L and R are selectively coupled to electrodes L and R, respectively.

The Hamiltonian of the DQWS, with the potential shown in Figure \ref{fig:dqw_energy_diagram}, 
is given by
\begin{alignat}{99}
	H_0 = \cfrac{\hat{p}_x^2 + \hat{p}_y^2}{2m} + H_\perp, 
\end{alignat}
where 
$\hat{p}_x$ and $\hat{p}_y$ are the momentum operators and
$m$ is the effective mass of the conduction band. 
The second term $H_\perp$, for the motion perpendicular to the DQWS, is given by
\begin{alignat}{99}
	H_\perp = - \cfrac{1}{2}\DSAS\htau_1 + \cfrac{\alpha}{\hbar}\htau_3 (\hat{p}_y\hsigx - \hat{p}_x\hsigy),
\end{alignat}
where the first term represents the interwell coupling and 
the second term expresses the antiparallel Rashba effective magnetic field.
Here 
$\htau_\gamma$ $(\gamma = 1, 2$ and $3)$ 
is the Pauli operator for pseudospin \cite{Ishikawa_Akera2019, Ishikawa_Akera2022, Kato2023} 
defined by
\begin{alignat}{99}
	\htau_1 &= \ket{\WR}\bra{\WL} + \ket{\WL}\bra{\WR},\\
	\htau_2 &= i\ket{\WR}\bra{\WL} - i\ket{\WL}\bra{\WR},\\
	\htau_3 &= \ket{\WL}\bra{\WL} - \ket{\WR}\bra{\WR},
\end{alignat}
where $\ket{\WL}$ and $\ket{\WR}$ represent the lowest bound state in the left and right wells, respectively. 

The eigenvector is given by $\Ket{n\sig\vk}=\Ket{n}\!\Ket{\sig}\!\Ket{\vk}$. 
Here $\Ket{\vk}$ is the eigenvector of $\hat{\vp} = (\hat{p}_x, \hat{p}_y)$ 
corresponding to the eigenvalue $\hbar\vk$. 
For each $\vk$, $\Ket{\sig}$ is defined by 
$\veb\cdot\hvsig\ket{\sig} = \sig \ket{\sig}$ 
where $\veb = k^{-1}(k_y, -k_x, 0)$ with $k=\sqrt{k_x^2+k_y^2}$ which is the unit vector in the direction of the effective magnetic field,
$\hvsig = (\hsigx, \hsigy, \hsigz)$, and $\sigma=\pm 1$. 
The vector $\Ket{n}$ ($n=\pm 1$) is given, for each $k$ and each $\sigma$, by 
\begin{alignat}{99}
	\Ket{n} &= \cfrac{1}{\sqrt{2}}\, \qty(\sqrt{1+n\sig\Delta_k} \Ket{\WL} -n \sqrt{1-n\sig\Delta_k} \Ket{\WR} ),\\
	\Delta_k &= 2\alpha k / \sqrt{\DSAS^2 + \qty(2\alpha k)^2}.
\end{alignat}
We also use $n=\WG$ for $n=-1$ and $n=\WE$ for $n=1$, 
since they are the ground state and the first excited state, respectively. 
The local spin polarization in each well of state $\Ket{n}$ at $\vk$ becomes  
\begin{alignat}{99}
	\sum_\sig\Braket{n\sig\vk|\hat{\vsigma}P_\xi|n\sig\vk} = \xi n \Delta_k\veb, 
\label{eq:local_spin_polarization}
\end{alignat}
where $P_\xi = \ket{\xi}\!\bra{\xi}$ is the projection operator onto well $\xi$ ($\xi=\WL, \WR$) 
with $\xi=1$ for $\xi=\WL$ and $\xi=-1$ for $\xi=\WR$.
Equation (\ref{eq:local_spin_polarization}) shows that 
the magnitude of the local spin polarization is $\Delta_k$, and 
its direction is opposite between $\xi=\WL$ and R and between $n=\WG$ and E 
as shown in Figure \ref{fig:dqw_energy_diagram}.

The eigenvalue of $H_0$ is given by
\begin{alignat}{99}
	\epsink &= \cfrac{\hbar^2k^2}{2m} +\cfrac{n}{2}\sqrt{\DSAS^2 + (2\alpha k)^2},
\label{eq:eigen_enegy}
\end{alignat}
The eigenvalue has no dependence on $\sig$ because the DQWS has the inversion symmetry. 

The Hamiltonian of electrode $\xi\ (=\WL, \WR)$ is assumed to be
\begin{alignat}{99}
	H_\mathrm{El} = \cfrac{\hat{p}_x^2+\hat{p}_y^2+\hat{p}_z^2}{2m_\mathrm{El}} + \epsi_{0} . 
\end{alignat}
Here the effective mass $m_\mathrm{El}$ is the same in $\xi =\WL \textrm{ and } \WR$, 
and $\epsi_0(<0)$ is the energy at the band bottom. 
The eigenvector is $\Ket{\xi \vk k_z\sigy}$ 
and the eigenenergy is 
\begin{alignat}{99}
	\epsi_{kk_z} = \cfrac{\hbar^2 (k^2 + k_z^2)}{2m_\mathrm{El}} + \epsi_{0} . \quad
\end{alignat}

\section{Boltzmann Equation and Distribution Function}
\label{sec:BoltzmannEquation}

We calculate the $\lTISP$, the spin current, and the inplane charge current 
of the DQWS model in Sec.~\ref{sec:DQWS_Model} by employing the Boltzmann equation in the first order of $\vnabla\Te$ 
and $\vnabla\muec = \vnabla\mu + e\vE$ 
where $\mu$ is the chemical potential, $e (> 0)$ is the absolute value of the electronic charge, and $\vE$ is the inplane electric field.
The distribution function $\fnk$ with $n$ the band index and $\vk$ the inplane wave vector is decomposed into the equilibrium distribution function
for local values of $\mu(\vecr)$ and $\Te(\vecr)$ with $\vecr =(x,y)$
and the deviation $\fone_{n\vk}$ in the first order of $\vnabla\Te$ and $\vnabla\muec$:
\begin{alignat}{99}
	\fnk = \fzero(\epsi_{n\vk}, \mu(\vecr), \Te(\vecr)) + \fone_{n\vk},   
	\label{eq:distribution_function}
\end{alignat}
where $\fzero(\epsi, \mu, \Te) = \qty{\exp[\qty(\epsi - \mu)/\kB \Te]+1}^{-1}$
with $\kB$ the Boltzmann constant.
The steady-state Boltzmann equation for $\fnk$ is given, 
in the relaxation time approximation with the momentum relaxation time $\taup$, by 
\begin{alignat}{99}
	\vvnk\cdot\cfrac{\partial \fnk}{\partial \vecr} + \cfrac{(-e)\vE}{\hbar}\cdot\cfrac{\partial \fnk}{\partial \vk}
	= - \cfrac{\fone_{n\vk}}{\taup}, 
	\label{eq:BoltzmannEq}
\end{alignat}
where $\vvnk = \hbar^{-1}\cfrac{\partial\epsink}{\partial\vk}$.
From Eq.(\ref{eq:BoltzmannEq}), $\fone_{n\vk}$ is obtained as \cite{Ashcroft_Mermin1976solid}
\begin{alignat}{99}
	\fone_{n\vk} = 
	 \cfrac{\taup}{\hbar}\cfrac{\partial \fzero}{\partial \vk}\cdot
	\left( \vnabla\muec + \cfrac{\epsink-\mu}{\Te}\vnabla\Te \right) .
	\label{eq:fone}
\end{alignat}

\section{Calculated Results}
\label{sec:Calculated_Results}
\subsection{Antiparallel TISP}
\label{sec:antiparallel_TISP}
The $\lTISP$ in well $\xi$ of the DQWS per unit area, $\vsigma_{\xi} = (\sigxxi, \sigyxi, \sigzxi)$, is 
given, using the first-order deviation $\fone_{n\vk}$, by 
\begin{alignat}{99}
	\vsigma_{\xi}
	&= \cfrac{1}{S}\sum_{n\sig\vk} \fone_{n\vk} \Braket{n\sig\vk|\hvsig P_\xi|n\sig\vk},
	\label{eq:localTISPvec}
\end{alignat}
with $S$ the system area.
Gradients $\vnabla\muec$ and $\vnabla\Te$ in $\fone_{n\vk}$ [Eq.(\ref{eq:fone})] are related 
by vanishing the charge current density, $\vjc$, which is given by
\begin{alignat}{99}
	\vjc
	&= -\cfrac{e}{S} \sum_{n\sig\vk} \fone_{n\vk} \vvnk
	= \Lmuec \vnabla \muec + \LT \vnabla \Te,
	\label{eq:jx_DQW}\\
	\Lmuec &= \sum_{n} \Lmuec_n, \quad  \Lmuec_n = \cfrac{2e\taup}{S\hbar}\sum_{\vk} \qty(\cfrac{k_x}{k})^2 \qty(-\cfrac{\partial\fzero}{\partial k}) v_{nk},
	\label{eq:Lmuec}\\
	\LT &= \sum_{n} \LT_n, \quad  \LT_n = \cfrac{2e\taup}{S\hbar}\sum_{\vk} \qty(\cfrac{k_x}{k})^2 \qty(-\cfrac{\partial\fzero}{\partial k}) \cfrac{\epsink-\mu}{\Te} v_{nk},
	\label{eq:LT}
\end{alignat}
where the dependence of $\taup$ on $n$ and $k$ is neglected, and $v_{nk}=|\vvnk|$.
From $\vjc=0$, we obtain $\vnabla\muec = -(\LT/\Lmuec)\vnabla\Te$.
We choose the $x$ axis in the direction of the temperature gradient, $\vnabla\Te=(\nablax\Te,0)$. 
Then the $\lTISP$ appears in the $y$ direction and $\sigyxi$ is given by 
\begin{alignat}{99}
	\sigyxi
	& = \xi\qty(\sigmuec \nabla_x \muec + \sigT \nabla_x \Te)
	= \xi\qty(-\sigmuec\cfrac{\LT}{\Lmuec} + \sigT) \nablax\Te,
	\label{eq:localTISP_DQW}\\
	\sigmuec &= \sum_{n} \sigmuec_{n}, \quad 
	\sigmuec_n = \cfrac{n\taup}{S\hbar}\sum_{\vk} \qty(\cfrac{k_x}{k})^2 \qty(-\cfrac{\partial\fzero}{\partial k}) \Delta_k,
	\label{eq:sigmuec}\\
	\sigT &= \sum_{n} \sigT_n, \quad 
	\sigT_n = \cfrac{n\taup}{S\hbar}\sum_{\vk} \qty(\cfrac{k_x}{k})^2 \qty(-\cfrac{\partial\fzero}{\partial k}) \cfrac{\epsink-\mu}{\Te} \Delta_k.
	\label{eq:sigT}
\end{alignat}
The $\lTISP$s in well L and R are in opposite direction, $\sigyL = -\sigyR$, 
and the contributions from subband G and E are also in opposite direction, 
$\sigmuec_{\rm G} = - \sigmuec_{\rm E}$ and $\sigT_{\rm G} = - \sigT_{\rm E}$. 

\iffigure
\begin{figure}
\centering
\begin{overpic}[scale=.5,]{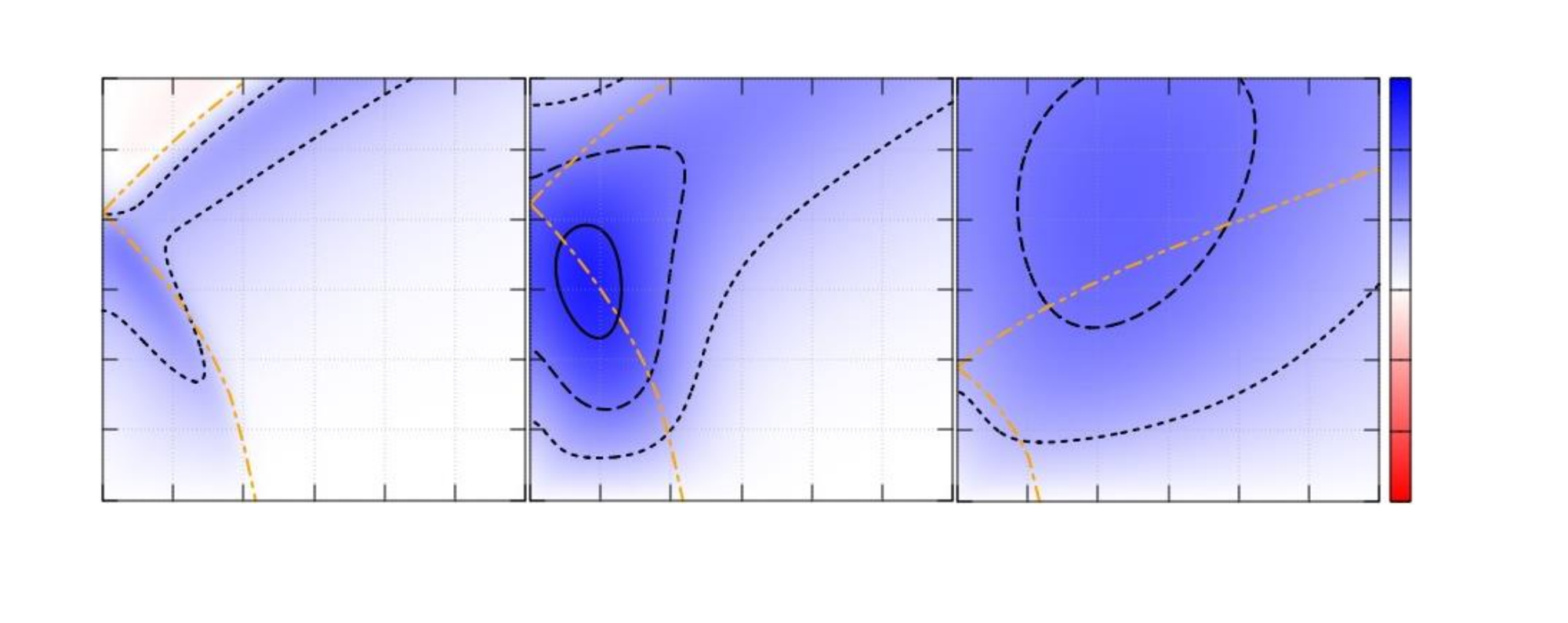}
	\put(	7,	37	){(a)$\tTe=0.04$}
	\put(	-0.5,	20.5	){$\talpha$}
	\put(	2.5,	12	){$0.5$}
	\put(	4.2,	16.5	){$1$}
	\put(	2.2,	20.5	){$1.5$}
	\put(	4.2,	25.5	){$2$}
	\put(	2.2,	29.4	){$2.5$}
	\put(	4.2,	34	){$3$}
	\put(	17,	2	){$\tDSAS$}
	\put(	4.2,	5	){$0$}
	\put(	9,	5	){$0.5$}
	\put(	15,	5	){$1$}
	\put(	18.3,	5	){$1.5$}
	\put(	24,	5	){$2$}
	\put(	27.3,	5	){$2.5$}
	\put(	32,	5	){$3$}
	\put(	18,	28	){$0.4$}
	\put(	34,	37	){(b)$\tTe=0.2$}
	\put(	45,	2	){$\tDSAS$}
	\put(	33.7,	5	){$0$}
	\put(	36.5,	5	){$0.5$}
	\put(	42,	5	){$1$}
	\put(	45.5,	5	){$1.5$}
	\put(	51,	5	){$2$}
	\put(	54.5,	5	){$2.5$}
	\put(	59,	5	){$3$}
	\put(	49,	23	){$0.4$}
	\put(	42,	27	){\textcolor{white}{$0.8$}}
	\put(	38,	23	){\textcolor{white}{$1.2$}}
	\put(	62,	37	){(c)$\tTe=1.0$}
	\put(	72,	2	){$\tDSAS$}
	\put(	61,	5	){$0$}
	\put(	64.5,	5	){$0.5$}
	\put(	69.2,	5	){$1$}
	\put(	73,	5	){$1.5$}
	\put(	78.5,	5	){$2$}
	\put(	82,	5	){$2.5$}
	\put(	87,	5	){$3$}
	\put(	82,	15	){$0.4$}
	\put(	76,	21	){\textcolor{white}{$0.8$}}
	\put(	96,	21	){$\tsig_{y\WL}$}
	\put(	90.5,	8	){$-1.5$}
	\put(	90.5,	12	){$-1$}
	\put(	90.5,	16.5	){$-0.5$}
	\put(	92.2,	20.7	){$0$}
	\put(	92.2,	24.7	){$0.5$}
	\put(	92.2,	29.3	){$1$}
	\put(	92.2,	34	){$1.5$}
\end{overpic}
\caption{\label{fig:TISP_alpha_DSAS}
The $\lTISP$ $\sigyL$ as functions of 
the temperature $\kB\Te$, 
the interwell-coupling strength $\DSAS$, 
and the Rashba-SOI intensity $\alpha$.
Dimensionless parameters are introduced as 
$\tsig_{y\WL} = \sigyL/\qty[\taup\kFzero/(4\pi\hbar)\nabla_x(\kB\Te)]$, $\tTe = \kB\Te/\epsiFzero$, 
$\tDSAS = \DSAS/\epsiFzero$, and $\talpha = 2\alpha\kFzero/\epsiFzero$.
$\epsiFzero$ and $\kFzero$ are the Fermi energy and the Fermi wave number at $\DSAS=\alpha=\Te=0$.
On the orange dashed-dotted (dashed double-dotted) line, 
the chemical potential coincides with the energy at $\vk=0$ of the first excited subband E (the ground subband G). 
}
\end{figure}
\fi

\figurename\ref{fig:TISP_alpha_DSAS} presents dependences of $\sigyL$ on 
the strength of the Rashba SOI ($\alpha$) and that of the interwell coupling ($\DSAS$).
In this calculation, we determine the chemical potential $\mu$ so that the electron number density 
\begin{alignat}{99}
	N_e
	&= \cfrac{1}{S}\sum_{n\sig\vk} \fnk
	\label{eq:N_electron}
\end{alignat}
is a constant. 
Then $\mu$ depends on values of $\alpha$, $\DSAS$, and $\Te$.
\figurename\ref{fig:TISP_alpha_DSAS}(a)--(c) show plots at different temperatures, $\tTe=\kB\Te/\epsiFzero=0.04, 0.2$, and 1.0, respectively,
where $\epsiFzero$ is the Fermi energy at $\DSAS=\alpha=\Te=0$.
The $\lTISP$ approaches zero at $\Te \rightarrow 0$ and at $\Te \rightarrow \infty$.
Our calculation shows that the largest value of $\sigyL$ appears around $\tTe\sim0.2$.
The orange dashed-dotted and dashed double-dotted lines correspond to 
the chemical potential coinciding with the energy at $\vk=0$ of the first excited subband E and that of the ground subband G, respectively 
(the bottom of the first excited subband is at $\vk=0$ for any values of $\DSAS$, 
while that of the ground subband is at $\vk=0$ for $\DSAS \ge 2\alpha\kso$ 
with $\kso=m\alpha/\hbar^2$).
We have found that larger values of the $\lTISP$ appear in the vicinity of these lines.  
This enhancement is most clearly seen in the panel of $\tTe=0.04$ 
and remains at a higher temperature of $\tTe=0.2$. 

\iffigure
\begin{figure}
\centering
\begin{overpic}[scale=.5,]{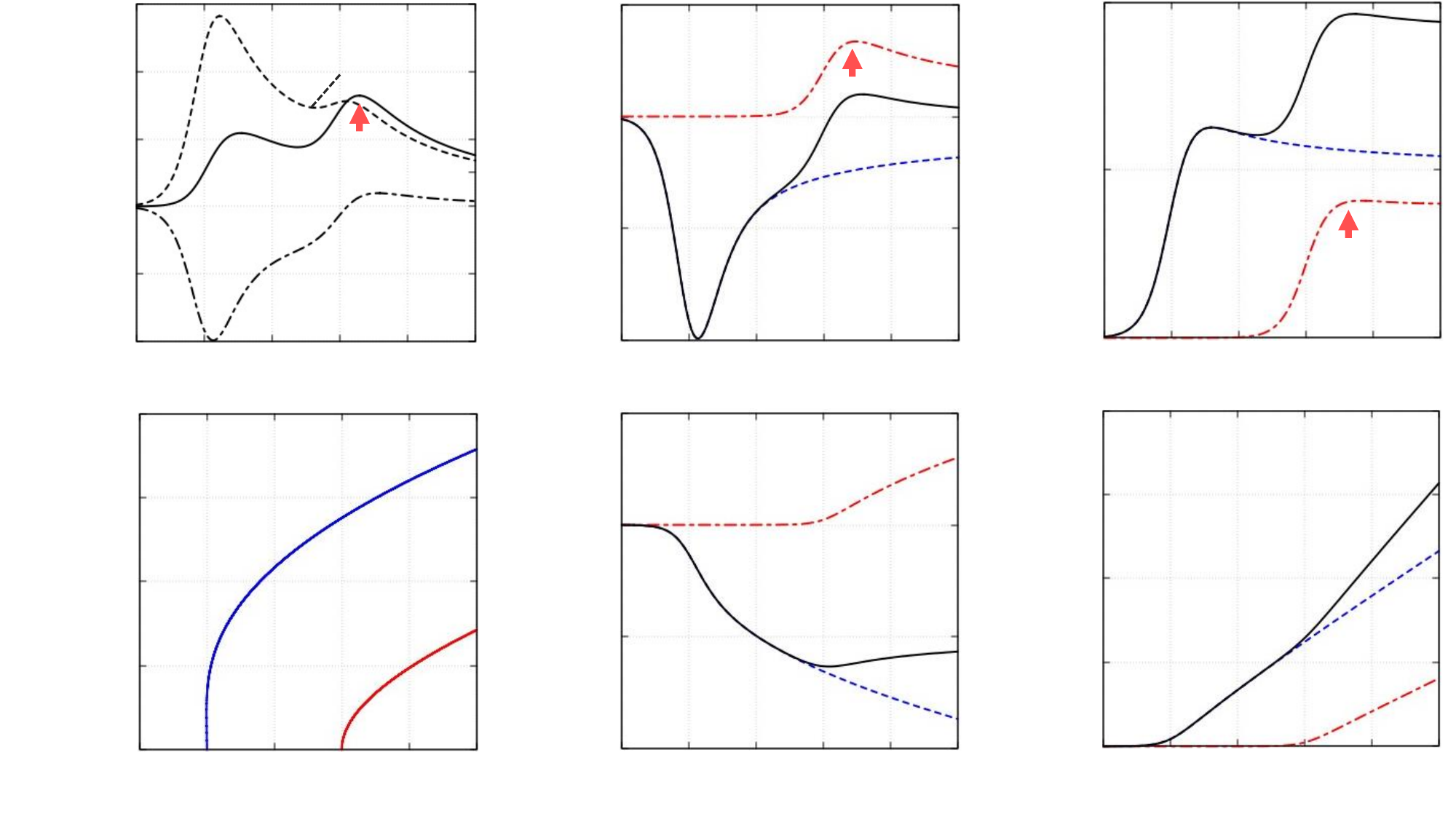}
	\put(	1,	55	){(a)}
	\put(	26,	40	){$\csigyLtwo$}
	\put(	24,	51.2	){$\csigyLone$}
	\put(	20,	44	){$\csigyL$}
	\put(	2,	44	){$\csigyL$}
	\put(	5.5,	32.8	){$-1$}
	\put(	3.5,	37	){$-0.5$}
	\put(	7.5,	41.5	){$0$}
	\put(	5.5,	46	){$0.5$}
	\put(	7.5,	51	){$1$}
	\put(	5.5,	55	){$1.5$}
	\put(	16,	28.5	){$\cmu-\cDSAS/2$}
	\put(	4.7,	30.8	){$-1.35$}
	\put(	10.8,	30.8	){$-0.9$}
	\put(	15.5,	30.8	){$-0.45$}
	\put(	22.9,	30.8	){$0$}
	\put(	26,	30.8	){$0.45$}
	\put(	31,	30.8	){$0.9$}
	\put(	22,	22	){\textcolor{blue}{$\WG$}}
	\put(	27,	12	){\textcolor{red}{$\WE$}}
	\put(	5,	15.5	){$\ck$}
	\put(	7.5,	5	){$0$}
	\put(	7.5,	10	){$1$}
	\put(	7.5,	15.5	){$2$}
	\put(	7.5,	21.5	){$3$}
	\put(	7.5,	27	){$4$}
	\put(	16,	0	){$\cepsi-\cDSAS/2$}
	\put(	4.7,	2.5	){$-1.35$}
	\put(	10.8,	2.5	){$-0.9$}
	\put(	15.5,	2.5	){$-0.45$}
	\put(	23,	2.5	){$0$}
	\put(	26,	2.5	){$0.45$}
	\put(	31,	2.5	){$0.9$}
	\put(	35,	55	){(b)}
	\put(	52,	45	){$\csigT$}
	\put(	57,	41	){\textcolor{blue}{$\csigTG$}}
	\put(	52,	50.5	){\textcolor{red}{$\csigTE$}}
	\put(	36,	44	){$\csigT$}
	\put(	39.2,	32.8	){$-1$}
	\put(	37.2,	39.5	){$-0.5$}
	\put(	40.4,	47	){$0$}
	\put(	39.2,	55	){$0.5$}
	\put(	49,	28.5	){$\cmu-\cDSAS/2$}
	\put(	38,	30.8	){$-1.35$}
	\put(	44,	30.8	){$-0.9$}
	\put(	48.5,	30.8	){$-0.45$}
	\put(	56,	30.8	){$0$}
	\put(	59,	30.8	){$0.45$}
	\put(	64.5,	30.8	){$0.9$}
	\put(	68,	55	){(c)}
	\put(	84.5,	50	){$\cLT$}
	\put(	91,	47	){\textcolor{blue}{$\cLTG$}}
	\put(	84,	36	){\textcolor{red}{$\cLTE$}}
	\put(	69,	44	){$\cLT$}
	\put(	74,	32.8	){$0$}
	\put(	72.5,	44	){$0.5$}
	\put(	74,	55	){$1$}
	\put(	82,	28.5	){$\cmu-\cDSAS/2$}
	\put(	71,	30.8	){$-1.35$}
	\put(	77,	30.8	){$-0.9$}
	\put(	82.2,	30.8	){$-0.45$}
	\put(	89.1,	30.8	){$0$}
	\put(	92,	30.8	){$0.45$}
	\put(	97.3,	30.8	){$0.9$}
	\put(	35,	27	){(d)}
	\put(	59,	12	){$\csigmuec$}
	\put(	57,	6.2	){\textcolor{blue}{$\csigmuecG$}}
	\put(	58,	19	){\textcolor{red}{$\csigmuecE$}}
	\put(	36,	15.3	){$\csigmuec$}
	\put(	38.7,	5	){$-4$}
	\put(	38.7,	12	){$-2$}
	\put(	40.7,	19	){$0$}
	\put(	40.7,	27	){$2$}
	\put(	49,	0	){$\cmu-\cDSAS/2$}
	\put(	38,	2.5	){$-1.35$}
	\put(	44,	2.5	){$-0.9$}
	\put(	48.5,	2.5	){$-0.45$}
	\put(	56,	2.5	){$0$}
	\put(	59,	2.5	){$0.45$}
	\put(	64.5,	2.5	){$0.9$}
	\put(	68,	27	){(e)}
	\put(	92,	20	){$\cLmuec$}
	\put(	93,	12	){\textcolor{blue}{$\cLmuecG$}}
	\put(	88.5,	7	){\textcolor{red}{$\cLmuecE$}}
	\put(	69,	15.3	){$\cLmuec$}
	\put(	74,	5	){$0$}
	\put(	74,	10	){$2$}
	\put(	74,	15.7	){$4$}
	\put(	74,	21.7	){$6$}
	\put(	74,	27	){$8$}
	\put(	82,	0	){$\cmu-\cDSAS/2$}
	\put(	71,	2.5	){$-1.35$}
	\put(	77,	2.5	){$-0.9$}
	\put(	82.2,	2.5	){$-0.45$}
	\put(	89.1,	2.5	){$0$}
	\put(	92,	2.5	){$0.45$}
	\put(	97.3,	2.5	){$0.9$}
\end{overpic}
\caption{\label{fig:TISP_cmu}
(a) 
The $\lTISP$ $\sig_{y\WL}$ 
and its decomposition into the first term $\sigyLone = (-\sigmuec\LT/\Lmuec)\nablax\Te$ 
and the second term $\sigyLtwo=\sigT\nablax\Te$ 
as a function of the chemical potential $\mu$. 
Dimensionless values are defined by 
$\csig_{y\WL}=\sig_{y\WL}/\sigma_0$, 
$\csigyLone=\sigyLone/\sigma_0$, 
$\csigyLtwo=\sigyLtwo/\sigma_0$, and
$\cmu=\mu/(2\alpha\kso)$  
with $\sigma_0=\taup\kso/(4\pi\hbar)\nabla_x(\kB\Te)$ and $\kso=m\alpha/\hbar^2$.  
$\DSAS$ and $\kB\Te$ are fixed at 
$\cDSAS=\DSAS/(2\alpha\kso)=0.9$ and $\cTe=\kB\Te/(2\alpha\kso) = 0.07$.
In the $\mu$ sweep, values of $\cDSAS$ and $\cTe$ do not change, 
while those of $\tDSAS$ and $\tTe$ (Figure \ref{fig:TISP_alpha_DSAS}) change.
The red arrow indicates the peak location of the $\lTISP$ around the bottom of the first excited subband.
The lower plot shows the ground (G) and the first excited (E) subbands with $\ck=k/\kso$.
(b)--(e)
Coefficients $\sigT, \LT, \sigmuec$, and $\Lmuec$,  
with their decompositions into the ground subband G and the first excited subband E,
as a function of the chemical potential.
$\csigmuec = \sigmuec/\sigalpha$ and 
$\csigT = \sigT/(\sigalpha\kB)$  
with $\sigalpha = \taup\kso/(4\pi\hbar)$,
while $\cLmuec=\Lmuec/\Lalpha$ and $\cLT=\LT/(\Lalpha\kB)$ 
with $\Lalpha=e\taup\alpha\kso/(\pi\hbar^2)$.  
}
\end{figure}
\fi

To closely investigate the above-mentioned enhancement around the orange lines [Figure \ref{fig:TISP_alpha_DSAS}], 
in Figure \ref{fig:TISP_cmu}(a) we plot $\sigyL$ 
as a function of the chemical potential $\mu$ at 
$(\cDSAS, \cTe) = (0.9, 0.07)$ where $\cDSAS=\DSAS/(2\alpha\kso)$ and $\cTe=\kB\Te/(2\alpha\kso)$.
The $\lTISP$ $\sigyL$ exhibits two peaks with changing $\mu$,   
one near the first excited subband bottom and 
the other near the ground subband bottom 
(the ground subband bottom is at $\vk=0$ when $\cDSAS \ge 1$).
Figure \ref{fig:TISP_cmu}(a) also shows that 
the first term $(-\sigmuec\LT/\Lmuec) \nablax\Te$ in Eq.(\ref{eq:localTISP_DQW}), 
which appears by imposing the zero-charge-current condition, 
is comparable in magnitude to 
the second term $\sigT \nabla_x \Te$. 
Signs of these two terms are opposite (same) in the vicinity of the ground (first excited) subband bottom.  
This is why the peak of $\sigyL$ near the first excited subband bottom is higher. 
To find out the origin of the peaks in $\sigyL$, 
we separately plot $\sigT$, $\sigmuec$, $\LT$, and $\Lmuec$  
in \figurename\ref{fig:TISP_cmu}(b)--(e). 
Then we find that $\sigT$ and $\LT$ with the factor $(\epsink-\mu)$ in their expressions, Eqs.(\ref{eq:sigT}) and (\ref{eq:LT}), 
exhibit larger variations around subband bottoms (leading to the peaks) compared to $\sigmuec$ and $\Lmuec$. 
\figurename\ref{fig:TISP_cmu}(b)--(e) also decompose 
each of $\sigT$, $\sigmuec$, $\LT$, and $\Lmuec$ into contributions from two subbands, 
confirming that the contribution of each subband shows a large variation around its bottom. 
Since $\sigmuec$ represents the local CISP, 
our calculation shows that the $\lTISP$ exhibits larger variations with the chemical potential 
in comparison to the local CISP. 

\iffigure
\begin{figure}
\centering
\begin{overpic}[scale=.51, ]{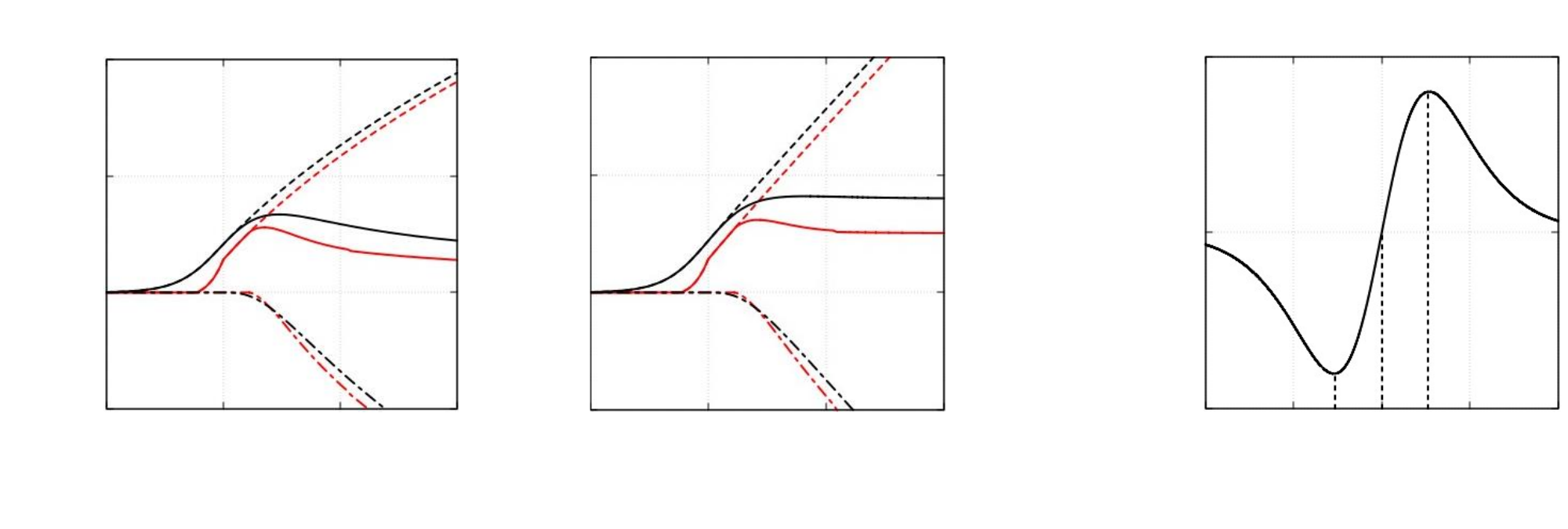}
	\put(	0,	30	){(a)}
	\put(	21.8,	20	){$\csigTE$}
	\put(	17,	24.8	){$\csigTe$}
	\put(	22.5,	10	){$\csigTh$}
	\put(	23.7,	14.2	){\textcolor{red}{$\csigTapp$}}
	\put(	24,	23.2	){\textcolor{red}{$\csigTeapp$}}
	\put(	13.5,	10	){\textcolor{red}{$\csigThapp$}}
	\put(	5,	28.7	){$1$}
	\put(	3,	21.7	){$0.5$}
	\put(	5.1,	14.5	){$0$}
	\put(	1.1,	7.5	){$-0.5$}
	\put(	13,	1.5	){$\cmu-\cDSAS/2$}
	\put(	3.5,	5	){$-0.45$}
	\put(	13.5,	5	){$0$}
	\put(	19.2,	5	){$0.45$}
	\put(	27.7,	5	){$0.9$}
	\put(	32,	30	){(b)}
	\put(	53,	22	){$\cLTE$}
	\put(	47,	25.5	){$\cLTe$}
	\put(	53.5,	9.5	){$\cLTh$}
	\put(	53,	16	){\textcolor{red}{$\cLTEapp$}}
	\put(	55,	25.5	){\textcolor{red}{$\cLTeapp$}}
	\put(	44.8,	9.5	){\textcolor{red}{$\cLThapp$}}
	\put(	36,	28.7	){$1$}
	\put(	34,	21.7	){$0.5$}
	\put(	36,	14.5	){$0$}
	\put(	32,	7.5	){$-0.5$}
	\put(	43.5,	1.5	){$\cmu-\cDSAS/2$}
	\put(	35,	5	){$-0.45$}
	\put(	44.8,	5	){$0$}
	\put(	50.8,	5	){$0.45$}
	\put(	59,	5	){$0.9$}
	\put(	70,	30	){(c)}
	\put(	74.5,	18	){0}
	\put(	62,	24	){$-\cfrac{\partial \fzero}{\partial\epsi}(\epsi-\mu)$}
	\put(	98,	5	){$\cepsi$}
	\put(	84.1,	5	){$\cepsih$}
	\put(	87.5,	5	){$\cmu$}
	\put(	90.5,	5	){$\cepsie$}
\end{overpic}
\caption{\label{fig:TISP_cmu_approx}
(a) The decomposition of $\csigTE$ at $(\cDSAS, \cTe)=(0.9, 0.07)$ in \figurename\ref{fig:TISP_cmu}(b) 
into the contribution from electron excitations ($\epsi-\mu\geq0$), $\sigTe$ [Equation (\ref{eq:sigTe})],  
and that from hole excitations ($\epsi-\mu<0$), $\sigTh$ [Equation (\ref{eq:sigTh})].  
Also plotted are approximate analytical expressions, 
$\csigTeapp$ [Equation (\ref{eq:sigTeapp})] and $\csigThapp$ [Equation (\ref{eq:sigThapp})],
with $\csigTapp=\csigTeapp + \csigThapp$.
(b) The same decomposition of $\cLTE$ in \figurename\ref{fig:TISP_cmu}(c). 
(c) The energy dependence of $[-\partial\fzero/\partial\epsi](\epsi-\mu)$ 
in expressions of $\sigTe$ and $\sigTh$ [Equations (\ref{eq:sigTe}) and (\ref{eq:sigTh})].
The maximum and the minimum are located around 
$\cepsie=\cmu+1.5\cTe$ and $\cepsih=\cmu-1.5\cTe$, respectively. 
}
\end{figure}
\fi

Now we focus on the higher peak of $\sigyL$ around the first excited subband bottom at $\epsi=\DSAS/2$, 
which mainly comes from the peak of $\sigTE$ and that of $\LTE$. 
In \figurename\ref{fig:TISP_cmu_approx}(a) 
$\sigTE$ is divided into the contribution from electron excitations ($\epsi-\mu\geq0$), $\sigTe$,  
and that from hole excitations ($\epsi-\mu<0$), $\sigTh$: 
\begin{alignat}{99}
	\sigTE &= \sigTe + \sigTh\\
	\sigTe
	&= \cfrac{\taup}{4\pi\hbar}\int^\infty_\frac{\DSAS}{2} \mathrm{d}\epsi \theta(\epsi-\mu) g(\epsi) \qty(-\cfrac{\partial \fzero}{\partial \epsi}) \cfrac{\epsi-\mu}{\Te}, 
	\label{eq:sigTe}\\
	\sigTh
	&= \cfrac{\taup}{4\pi\hbar}\int^\infty_\frac{\DSAS}{2} \mathrm{d}\epsi \theta(-\epsi+\mu) g(\epsi) \qty(-\cfrac{\partial \fzero}{\partial \epsi}) \cfrac{\epsi-\mu}{\Te}
	\label{eq:sigTh}
\end{alignat}
where $g(\epsi)=k\Delta_k$ 
with $k$ determined by $\epsi=\epsink$ [Equation (\ref{eq:eigen_enegy})]
and $\theta(\epsi)$ is defined by $\theta(\epsi)=1\ (\epsi\geq0)$ and $\theta(\epsi)=0\ (\epsi<0) $.
In the chemical-potential region below the subband bottom ($\mu < \DSAS/2$) 
the hole contribution $\sigTh$ is absent and 
the positive electron contribution $\sigTe$ increases with increasing the chemical potential.
On the other hand, in the region of $\mu \geq \DSAS/2$, 
the negative hole contribution $\sigTh$ starts to suppress $\sigTE$ which eventually decreases with increasing $\mu$.  
Thus the peak of $\sigTE$ appears near the subband bottom.  

The same explanation holds for the peak of $\LTE$. 
\figurename\ref{fig:TISP_cmu_approx}(b) presents chemical-potential dependences of 
its electron and hole contributions, $\LTe$ and $\LTh$, given by
\begin{alignat}{99}
	\LTE &= \LTe + \LTh,\\
	\LTe
	&= \cfrac{e\taup}{2\pi\hbar}\int^\infty_\frac{\DSAS}{2} \mathrm{d}\epsi \theta(\epsi-\mu)h(\epsi) \qty(-\cfrac{\partial \fzero}{\partial \epsi}) \cfrac{\epsi-\mu}{\Te}, 
	\label{eq:LTe}\\
	\LTh
	&= \cfrac{e\taup}{2\pi\hbar}\int^{\infty}_\frac{\DSAS}{2} \mathrm{d}\epsi \theta(-\epsi+\mu)h(\epsi) \qty(-\cfrac{\partial \fzero}{\partial \epsi}) \cfrac{\epsi-\mu}{\Te},
	\label{eq:LTh}
\end{alignat}
where $h(\epsi)=kv_{\WE k}$.
Therefore the change in the balance between electron and hole contributions in $\sigTE$ and $\LTE$ 
is responsible for the formation of their peaks, leading to the peak formation in the $\lTISP$ $\sigyL$.

Finally we consider 
the reason why the maximum of $\lTISP$ $\sigyL$ appears at a non-zero interwell coupling $\DSAS$ in \figurename\ref{fig:TISP_alpha_DSAS}
by estimating the $\DSAS$ dependence of the peak height in each of $\sigTE$ and $\LTE$ 
with use of their approximate analytical expressions.
The approximation is made 
by replacing $g(\epsi)$ in Eq.~(\ref{eq:sigTe}) [Eq.~(\ref{eq:sigTh})] 
by its value at $\epsie = \mu+1.5\kB\Te$ [$\epsih = \mu-1.5\kB\Te$] 
because at $\epsi=\epsie$ [$\epsi=\epsih$] 
the absolute value of $\qty[-\partial \fzero/\partial \epsi](\epsi-\mu)$ in Eq.~(\ref{eq:sigTe}) [Eq.~(\ref{eq:sigTh})] 
is nearly the largest 
in $\epsi>\mu$ [$\epsi<\mu$] as shown in \figurename\ref{fig:TISP_cmu_approx}(c). 
Then the approximate expressions become 
\begin{alignat}{99}
	\sigTeapp
	&= \cfrac{\taup}{4\pi\hbar}g(\mu+1.5\kB\Te)\int^\infty_\frac{\DSAS}{2} \mathrm{d}\epsi \theta(\epsi-\mu) \qty(-\cfrac{\partial \fzero}{\partial \epsi}) \cfrac{\epsi-\mu}{\Te}, 
	\label{eq:sigTeapp}\\
	\sigThapp
	&= \cfrac{\taup}{4\pi\hbar}g(\mu-1.5\kB\Te)\int^\infty_\frac{\DSAS}{2} \mathrm{d}\epsi \theta(-\epsi+\mu) \qty(-\cfrac{\partial \fzero}{\partial \epsi}) \cfrac{\epsi-\mu}{\Te}.
	\label{eq:sigThapp}
\end{alignat}
Similarly approximate expressions, $\LTeapp$ and $\LThapp$, are obtained. 
\figurename\ref{fig:TISP_cmu_approx}(a) and (b) show that 
thus obtained approximate estimates $\sigTEapp=\sigTeapp+\sigThapp$ and $\LTEapp=\LTeapp+\LThapp$ (red lines) 
well reproduce $\sigTE$ and $\LTE$ (black lines).
Since the peak in each of $\sigTE$ and $\LTE$ is located around $\mu=\DSAS/2+1.5\kB\Te$, 
we estimate the peak height at $\mu=\DSAS/2+1.5\kB\Te$ where the hole contribution $\sigThapp=\LThapp=0$. 
Then estimated peak values $\sigTEapppeak$ and $\LTEapppeak$ become 
\begin{alignat}{99}
	\sigTEapppeak &= \cfrac{\taup\kB\ln2}{4\pi\hbar}g\qty(\cfrac{\DSAS}{2}+3\kB\Te),
	\label{eq:max_sigTEapp}\\
	\LTEapppeak &= \cfrac{e\taup\kB\ln2}{2\pi\hbar}h\qty(\cfrac{\DSAS}{2}+3\kB\Te).
	\label{eq:max_LTEapp}
\end{alignat}
Both $\sigTEapppeak$ and $\LTEapppeak$ increase by switching on the interwell coupling,  
that is
$\frac{d}{d\DSAS}\sigTEapppeak \!>\!0$ and 
$\frac{d}{d\DSAS}\LTEapppeak \!>\!0$ at $\DSAS=0$, 
which are derived from 
$\frac{d}{d\DSAS}k>0$,
$\frac{d}{d\DSAS}\Delta_k=0$, and
$\frac{d}{d\DSAS}v_{\WE k}>0$ at $\DSAS=0$.
Here $k$ is determined by $\epsiEk = \DSAS/2+3\kB\Te$ and then depends on $\DSAS$. 
Therefore enhancements of the momentum ($k$) and the velocity ($v_{\WE k}$) by $\DSAS$ at $\DSAS=0$ 
lead to those of $\sigTEapppeak$ and $\LTEapppeak$,  
which can explain the reason 
why the $\lTISP$ exhibits its maximum at a non-zero $\DSAS$ in \figurename\ref{fig:TISP_alpha_DSAS}.

\subsection{Spin Current}
\label{sec:spin_current}

In this section we show that the spin current into the adjacent electrode is proportional to the $\lTISP$ 
as derived for the spin current extracted from the local CISP \cite{Suzuki2023} 
in a simplified model of the electron tunneling to the electrode.
The spin current per unit area, $\js_{z,\xi}$, from the DQWS to electrode $\xi$ is given by 
\begin{alignat}{99}
	\js_{z,\xi}
	=	\cfrac{1}{S}\sum_{\sigy}\cfrac{\hbar\sigy}{2}\sum_{n\sig\vk}\sum_{k_z} W_{\xi\vk k_z\sigy,n\sig\vk} [\fnk - \fzero(\epsi_{\vk\kz}, \mu^\mathrm{eq}, \Te^\mathrm{eq})], 
	\label{eq:js_DQW}
\end{alignat}
where we assumed that each electrode is in equilibrium with the chemical potential $\mu^{\rm eq}$ and the temperature $\Te^{\rm eq}$,
which are spatially uniform in contrast to spatially-varying $\mu(\vecr)$ and $\Te(\vecr)$ 
in $\fzero$ [Equation (\ref{eq:distribution_function})] of the DQWS.
The tunneling rate between DQWS and electrode $\xi$ 
is given by
\begin{alignat}{99}
	W_{\xi\vk\kz\sigy,n\vk\sig} = \cfrac{2\pi}{\hbar}\qty|\Braket{\xi\vk\kz\sigy| H_\mathrm{T} |n\sig\vk}|^2 \delta(\epsi_{nk} - \epsi_{k\kz}).
	\label{eq:Fermis_golden_rule}
\end{alignat}
We assume that the tunneling occurs between the well and the electrode in the same side 
without changing the inplane momentum $\vk$ and the spin. 
Then the tunneling Hamiltonian $H_\mathrm{T}$ is given by
\begin{alignat}{99}
	H_\mathrm{T} =\sum_{\xi \vk k_z\sigy} 	
		  \Ket{\xi\vk\sigy} 
		  \Braket{\xi\vk\sigy|H_\mathrm{T}|\xi\vk k_z\sigy} 
		  \Bra{\xi\vk k_z \sigy} +\textrm{h.c.} ,
	\label{eq:Tunneling_Hamiltonian}
\end{alignat}
where $\Ket{\xi\vk\sigy}=\Ket{\xi}\!\Ket{\vk}\!\Ket{\sigy}$ is the wave function localized in well $\xi$ 
and h.c.~denotes the Hermitian conjugate of the preceding term. 
We additionally assume that the matrix element $\Braket{\xi\vk\sigy|H_\mathrm{T}|\xi\vk k_z\sigy}$ 
has no dependence on $\vk$, $\sigy$, and $\xi$. 
The distribution difference in Equation (\ref{eq:js_DQW}) is divided into 
$\fnk - \fzero(\epsi_{n\vk}, \mu(\vecr), \Te(\vecr))=\fone_{n\vk}$ and 
$\fzero(\epsi_{n\vk}, \mu(\vecr), \Te(\vecr)) - \fzero(\epsi_{\vk\kz}, \mu^\mathrm{eq}, \Te^\mathrm{eq})$.  
The latter produces no spin current because it is isotropic in $\vk$ space and 
the former $\fone_{n\vk}$ gives 
\begin{alignat}{99}
	\js_{z,\xi}
	&= \cfrac{\hbar}{2S} \sum_{n\sig\vk}
	\cfrac{1}{\tau_\xi} \fone_{n\vk}
	\Braket{n\sig\vk| \hsigy P_\xi |n\sig\vk}
\end{alignat}
where $\tau_{\xi}$ is the lifetime of state $\ket{\xi\vk\sigy}$ due to tunneling into electrode $\xi$,  
defined by
\begin{alignat}{99}
	\cfrac{1}{\tau_\xi}
	&= \sum_{k_z} \cfrac{2\pi}{\hbar}\qty|\Braket{\xi\vk\sigy|H_\mathrm{T}|\xi\vk k_z\sigy}|^2 
	\delta(\epsi_{n\vk} - \epsi_{\vk k_z}).
	\label{eq:transition_probability_of_electrode}
\end{alignat}
We neglect the variation of $\Braket{\xi\vk\sigy|H_\mathrm{T}|\xi\vk k_z\sigy}$ with $k_z$ 
by choosing a sufficiently large value of $|\epsi_0|$ 
such that $|\epsi_0| \gg \epsiFzero$ \cite{Suzuki2023}.
Then the spin current is shown to be proportional to the $\lTISP$ $\sig_{y\xi}$: 
\begin{alignat}{99}
	\js_{z,\xi}
	= \cfrac{1}{\tau_\xi} \cfrac{\hbar}{2} \sig_{y\xi}.
\end{alignat}

\section{Conclusions} 
\label{sec:Conclusions}
We have considered the double-quantum-well structure (DQWS) as the simplest system with locally-broken inversion symmetry 
and theoretically studied thermally-induced local spin polarization ($\lTISP$)  
which can be extracted as spin current. 
We have calculated the $\lTISP$ using the Boltzmann equation in the relaxation-time approximation 
under the condition of zero charge current 
with an aim to search optimum strengths of the Rashba SOI and the interwell coupling 
which maximize the magnitude of the $\lTISP$. 
We have found that the $\lTISP$ exhibits the maximum at a finite Rashba SOI and a nonzero interwell coupling 
when the electron density is fixed. 
The finite optimum value of the Rashba-SOI stems from the fact that 
the optimum position of the chemical potential to maximize the $\lTISP$ is near the first-excited subband bottom 
where the thermal current transfers electrons from the ground subband with negative local spin polarization 
to the first-excited subband with positive local spin polarization, 
leading to the largest induced spin polarization. 
We have also presented a possible explanation for the nonzero optimum value of the interwell coupling: 
the interwell coupling increases magnitudes of the momentum and the group velocity at the most probable energy 
in the first-excited subband 
and consequently enhances the $\lTISP$. 
We have derived the formula for the spin current into an electrode, 
which is generated from the antiparallel TISP with a selective coupling of the electrode to one well of the DQWS, 
and shown that it is proportional to the $\lTISP$ in the well 
by assuming the same tunneling rate to the electrode for all occupied eigenstates in the DQWS.

We have also confirmed, from Figure \ref{fig:TISP_alpha_DSAS} (a) and (b) (along $\DSAS=0$ line), 
that the largest TISP 
in a decoupled well, that is the 2DES, 
appears at the chemical potential near the bottom of the first-excited subband at $\vk=0$ 
where the ground and first-excited subbands have the same energy in the 2DES. 
The TISP of the 2DES with the Rashba SOI was already calculated in several previous studies 
\cite{Wang2010, Dyrdal2013, Tolle2014, Xiao2016, Dyrdal2018}. 
Some of them \cite{Dyrdal2013, Xiao2016, Dyrdal2018} calculated the TISP 
without imposing the condition of zero charge current, 
that is in the absence of the electrochemical-potential gradient 
and showed \cite{Dyrdal2013, Dyrdal2018} 
that the largest absolute value of the TISP appears near the ground subband bottom 
(which is at nonzero $\vk$ in the 2DES). 
In this paper we have found, in the standard experimental condition with zero charge current, 
that the higher peak appears near the first-excited subband bottom 
because the peak in the ground subband is suppressed by the opposite contribution from the electrochemical-potential gradient. 
Other calculations \cite{Wang2010, Tolle2014} of the TISP in the 2DES 
were made under the condition of zero charge current. 
These calculations focused on the region of the Fermi energy much larger than $\kB\Te$
which is away from the first-excited subband bottom where we have found the maximum of the TISP.

In this paper we have revealed that the $\lTISP$ (the TISP) is the largest 
at the chemical potential near the bottom of the first-excited subband in the DQWS (2DES) 
and that the coupling between wells in the DQWS enhances the $\lTISP$.  
These findings will be useful in designing the optimum TISP device.  

\section*{Acknowledgment} 

This work was partly supported by 
Grant-in-Aid for Scientific Research (C) Grant No. JP21K03413
from the Japan Society for the Promotion of Science (JSPS).

\bibliography{Spintronics_v2_ysuzuki}

\end{document}